\documentclass[amsmath,amssymb,nofootinbib]{revtex4-2}
\usepackage{graphicx}
\usepackage{dcolumn}
\usepackage{color}
\usepackage{bm}
\usepackage{amsmath,amssymb,graphicx,hyperref,xcolor}
\usepackage{epsfig}
\usepackage{enumerate}
\usepackage{array}
\usepackage{multirow}
\usepackage{graphicx} 
\usepackage{subcaption}
\usepackage{soul,xparse}
\usepackage[section]{placeins}

\begin{document}
\title
{Painlev\'{e}-Gullstrand coordinates for Kiselev black holes}
\author{Bijan Bagchi\footnote{E-mail: bbagchi123@gmail.com}$^1$, 
 Sauvik Sen\footnote{E-mail: sauviksen.physics@gmail.com}$^2$}

\vspace{2mm}

\affiliation{$^{1}$Department of Applied Mathematics, University of Calcutta, Kolkata 700009, India\\
$^{2}$ Department of Physics, Shiv Nadar Institution of Eminence, Gautam Buddha Nagar, Uttar Pradesh 203207, India}

\vskip-2.8cm
\date{\today}
\vskip-0.9cm

\begin{abstract}

We investigate the implications provided by the modified Painlev\'{e}-Gullstrand coordinates in the context of quintessence for the Kiselev black hole. In this regard, we set up a fully static line element in terms of lapse and shift functions, apart from including the deformation parameter signaling deviation from the standard Painlev\'{e}-Gullstrand metric. We address two specific issues pertaining to the problems of radiation and dust furnished by the corresponding barotropic index parameter and study the related consequences by performing a range of analyses to explore the influence imposed by quintessence. We also discuss the thermodynamical consequences by evaluating the expressions of the Hawking temperature and the entropy function in closed forms.

\end{abstract}

\maketitle

\section{Introduction}

The term \textit{quintessence} was coined in a 1998 paper by Caldwell et al \cite{cald} after interest in it picked up (see, for reviews, \cite{shinji, andr}) following influential works by Ratra and Peebles \cite{ratra1} (see also \cite{ratra2}), and Wetterich \cite{wett}. The core requirement of quintessence is to understand the accelerated rate of the expansion of the universe, where it plays the central role of a hypothetical form of dark energy. Toward this end, one introduces a minimally coupled scalar field to gravity in which the cosmic fluid is assumed to be dynamic \cite{zlat, wein, noza1}. The scalar field is beset with a spatially inhomogeneous negative pressure component that changes over time, accompanied by a gradual evolution of the energy function whose slowness has the implication that the associated potential energy density exceeds its kinetic energy counterpart: the nature of quintessence being attractive or repulsive depends on what the ratio of its kinetic and potential energy signifies. Among the different plausible outcomes of quintessence \cite{sund}, Kiselev \cite{kise} determined the density of quintessence matter on a Schwarzschild background, while thermodynamical aspects were studied in \cite{hamil}. Note that quintessence is not to be confused with the non-changing cosmological constant, since to afford any explanation of dark energy, it has to be dynamic in character. \\  

The purpose here is to examine quintessence in an extended framework of the Painlev\'{e}-Gullstrand coordinates based on the Arnowitt-Deser-Misner (ADM) formalism for the Kiselev black hole (see, for instance, \cite{hendi, sade, jiao} and references therein). Previously, such a class of spherically symmetric black holes was explored as the one with short hair \cite{vern}. In the realm of non-linear geometry, subsequent studies have been carried out \cite{dari1, dari2}. Moreover, the inclusion of the effects of higher-order WKB expansions has been estimated \cite{hasnain}. In the following, our strategy is to follow the procedure described in \cite{volovik} where it was shown that the ADM procedure is typically useful for the Hamiltonian formulation formulated in terms of Poisson brackets. \\

Let us focus on the following modification of Painlev\'{e}-Gullstrand coordinates by focusing on an underlying metric given in general terms

\begin{equation}
    ds^2 = \left (\mathcal{N}^2 - \alpha_{ik} \mathcal{N}^i\mathcal{N}^k  \right )dt^2 -2 \alpha_{ik}\mathcal{N}^k dt dx^i -\alpha_{ik}dx^idx^k
    \label{ADM1}
\end{equation}
where $\mathcal{N}$ is the lapse function, $\mathcal{N}^i$ is the shift function, and $\alpha_{ik}$ denote the space components of the ADM metric. The spherically symmetric ADM line element following (\ref{ADM1}) is similar to the acoustic metric, where $\mathcal{N}^i$ plays the role of the flow velocity. With $\mathcal{N}^i = v^i$, one can interpret it in the form

\begin{equation}
    ds^2 = \mathcal{N}^2 dt^2 - \alpha_{rr} \left (dr - v dt \right )^2 - r^2 d\theta^2 - r^2 \sin^2\theta\,d\phi^2
\end{equation}
The crossed terms can be removed with the help of the transformation $dt \rightarrow d\bar{t} -g(r) dr$, which leaves one with 

\begin{equation}
    ds^2 =  (\mathcal{N}^2 - v^2 \alpha_{rr}) d\bar{t}^2 - \alpha_{rr} (\mathcal{N}^2 - v^2 \alpha_{rr})^{-1} \mathcal{N}^2 dr^2 - r^2 d\Omega^2
\end{equation}
where $r^2 d\Omega^2 = r^2 d\theta^2 + r^2 \sin^2\theta\,d\phi^2$, and $g(r)$ has the restriction

\begin{equation}
    g(r) = \frac{\alpha_{rr} v}{\mathcal{N}^2 - v^2 \alpha_{rr}}
\end{equation}
With the choice $\alpha_{rr} = \frac{1}{\mathcal{N}^2}$, one recovers the fully static line-element 

\begin{equation}
      ds^2 = f(r) d\bar{t}^2 - f(r)^{-1} dr^2 - r^2 d\Omega^2
\end{equation}
accompanied by the pair 

\begin{equation}\label{f(r)}
       f(r) = \mathcal{\mathcal{N}}^2-\frac{v^2}{\mathcal{N}^2}, \quad g(r) = \frac{v(r)}{\mathcal{N}^4(r) -v^2 (r)}
\end{equation}
The above steps serve as the underlying equations for the modified Painlev\'{e}-Gullstrand scenario, which corresponds to the case $\mathcal{N} \neq 1$. Of course, the conventional set of Painlev\'{e}-Gullstrand coordinates holds for $\mathcal{N} = 1$ when 

\begin{equation}
    ds^2 = dt^2 - \left (dr -v(r) dt \right )^2 - r^2 d\Omega^2
\end{equation}
Note that we have identified the shift function as the velocity vector field of the freely falling observer, thus relating the spatial coordinate changes between these foliated hypersurfaces in terms of the ADM formalism.\\

In \cite{volovik}, an attempt was made to extend the Painlev\'{e}-Gullstrand coordinates by adopting a phenomenological form for the shift function $v(r)$ when the lapse function deviated from the unity value. This allowed one to deal with the Schwarzschild-de Sitter black hole, which is equipped with two horizons in the ADM framework. 
In the following, our task will be to investigate the modified scheme of Painlev\'{e}-Gullstrand coordinates for the Kiselev black hole in the background environment embedded by radiation and dust by exploring similar techniques. We will study these two cases for different variants of quintessence, as indicated by the specific values of the barotropic index parameter. Specifically, the existence of the stationary character of the shift function will be our point of focus whose role, as we will find, is qualitatively different from the Schwarzschild-de Sitter spacetime. 
Throughout our work, we shall adopt units $c = G = \hbar = k_B = 4\pi\varepsilon_0 = 1$, where $c$ is the speed of light in vacuum, $G$ is the gravitational constant, $\hbar$ is the reduced Planck constant, $k_B$ is the Boltzmann constant, and $\varepsilon_0$ is the permittivity of free space. 

\section{Modified Painlev\'{e}-Gullstrand coordinates and quintessence}

The Kiselev's model of an electrically charged black hole corresponds to the metric function $f(r)$ which is of the form \cite{kise, sade}
\begin{equation}\label{f^q(r)}
    f (r) = 1-\frac{2M}{r} + \frac{Q^2}{r^2}-\frac{\alpha}{r^{3\omega+1}},
\end{equation}
where $M$ is the ADM mass, $Q$ is the electric charge, $\alpha >0$ is a normalization parameter that indicates the intensity of the quintessence field, and the barotropic index parameter $\omega$ reflects the nature of the quintessence field. The role of the parameter $\omega$ is to account for the late-time cosmic acceleration. The specific values $\omega = \frac{1}{3}$ and $\omega = 0$ correspond to whether the black hole is surrounded by radiation or dust, respectively.  Kiselev's model was an attempt\footnote{However, the idea of quintessence does not go without problems \cite{seshadri}. For one thing, there is the fine-tuning problem concerning the smallness of the energy density in relation to other scales involved, while for another, there is the cosmic coincidence problem, that requires setting up of the initial condition highly precisely such that the missing energy density and matter density become comparable today. Furthermore, as pointed out in \cite{visser}, the Kislev black hole spacetime does not represent a perfect fluid spacetime.} to solve Einstein's equations for quintessence matter and determine solutions in terms of $\omega$ and $\alpha$.  We will discuss the two cases of $\omega$ separately. We first consider the case of radiation.\\

\subsection{The case of radiation}

Here $f(r)$ is given by ($\omega = \frac{1}{3}$) \cite{sade}

\begin{equation}
     f(r) = 1-\frac{2M}{r} + \frac{Q^2-\alpha}{r^2}
     \label{f_rad_equation}
\end{equation}

\begin{figure}[!h]
\centering

\begin{subfigure}{0.45\textwidth}
    \centering
    \includegraphics[width=\linewidth]{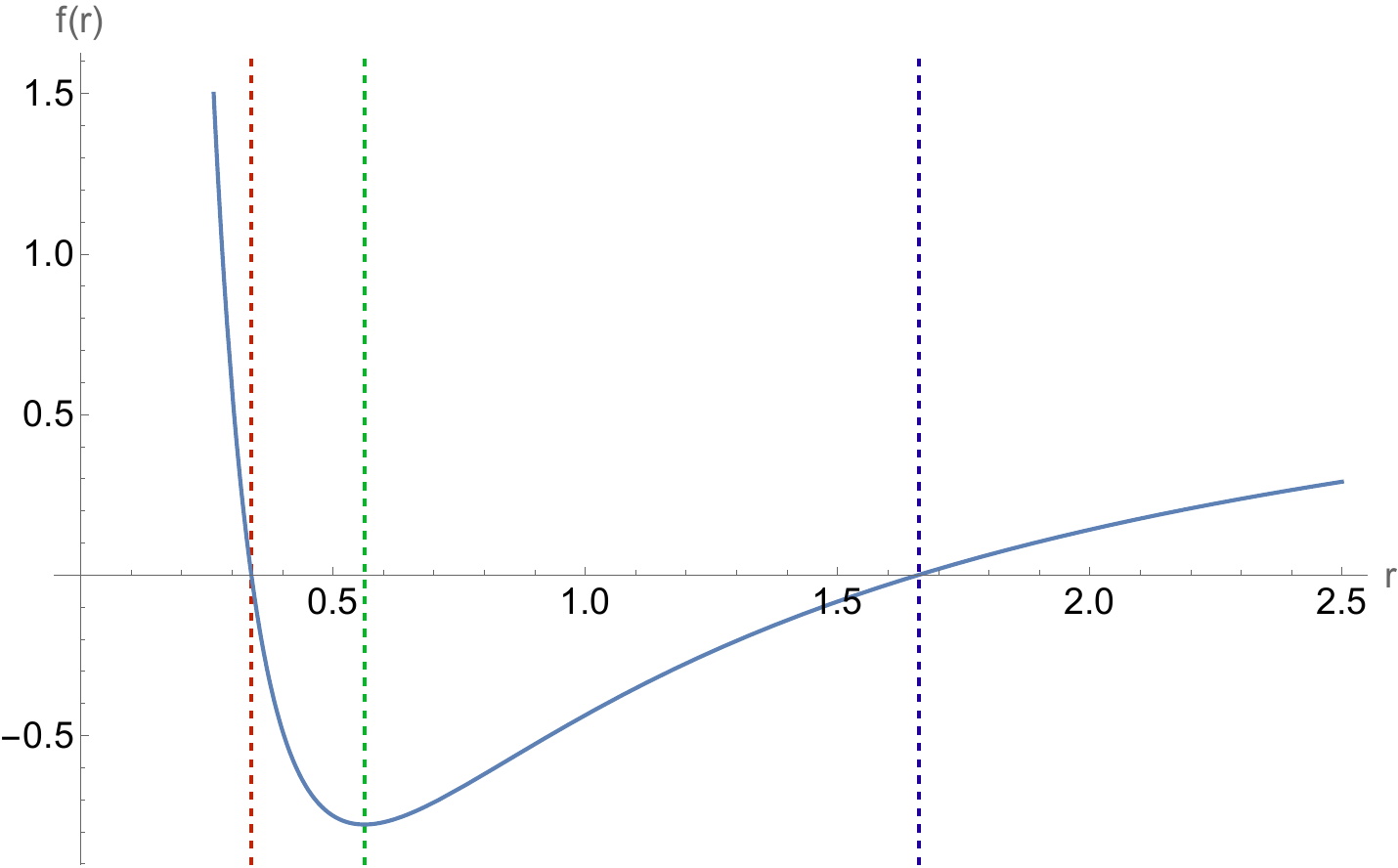}
        \caption{$\alpha=0$}
\end{subfigure}
\hfill
\begin{subfigure}{0.45\textwidth}
    \centering
    \includegraphics[width=\linewidth]{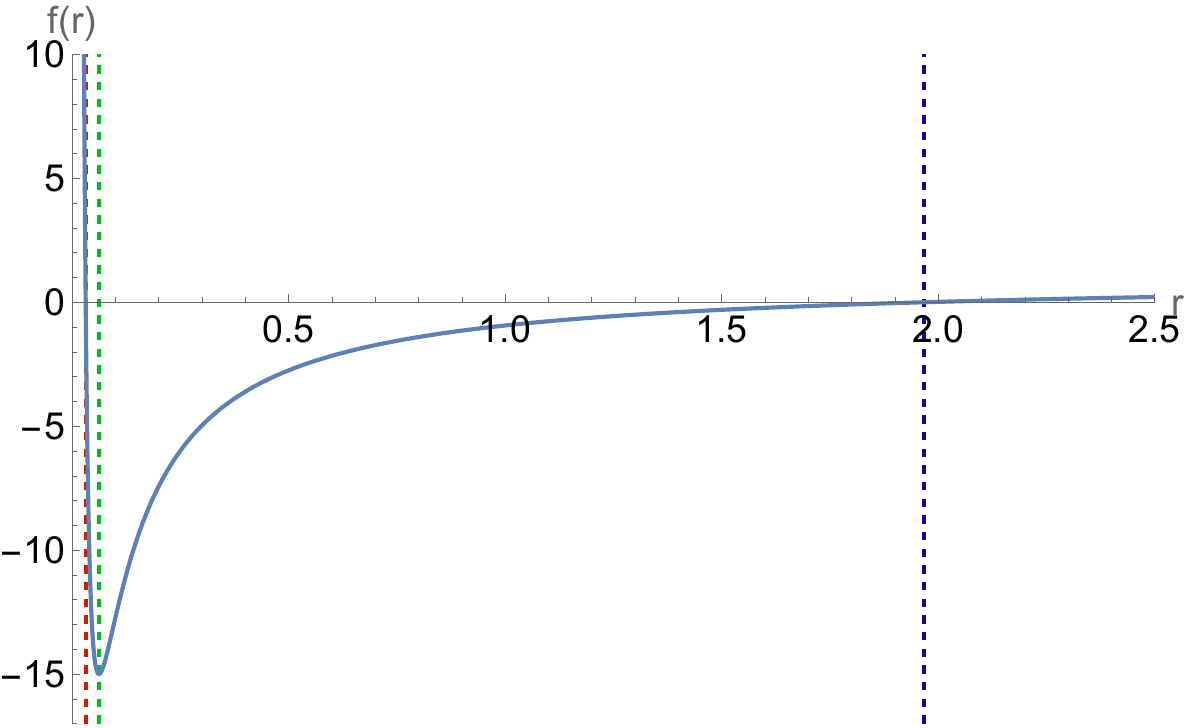}
        \caption{$\alpha=0.5$}
\end{subfigure}

\caption{Variation of $f(r)$ with $r$ for different values of $\alpha$ for the radiation case. The blue dashed line represents the outer horizon, the red one represents the inner horizon, and the green dotted line corresponds to where $r_0$ is minimum. We have taken $M=1$ and $Q=0.75$ to exhibit the sub-extremal condition.}
\label{f(r)_rad}
\end{figure}

In the conventional Painlev'e-Gullstrand scheme ($\mathcal{N}=1$), the accompanying shift function is given by the reduced relation 

\begin{equation}
     v^2 (r) = \frac{2M}{r} - \frac{Q^2-\alpha}{r^2}
\end{equation}
Horizon limits are
calculated from (\ref{f_rad_equation}) by setting $f (r) = 0$. In this way, we determine the outer limit ($r^{rad}_{+}$) and the inner limit ($r^{rad}_{-}$) to be 

\begin{equation}
     r^{rad}_{\pm} = M \pm \sqrt{M^2 - (Q^2-\alpha)}
     \label{horizon_rad}
\end{equation}
The condition of the reality of the square root terms in (\ref{horizon_rad}) leads to $M \geq (Q^2-\alpha)$ and $|Q| \geq \alpha$. 
For such a region, we have the criterion 

\begin{equation}
    r < \frac{r^{rad}_+r^{rad}_-}{r^{rad}_+ + r^{rad}_-} = \frac{Q^2-\alpha}{2M}
\end{equation}

In Fig.\ref{f(r)_rad} we plotted the behavior of $f(r)$ as a function of $r$ for two different values of $\alpha$. We note that $f(r)$ decreases below a certain value of $r$ by descending to a minimum position and then increases monotonically after undergoing a sign change. The minimum value of $f(r)$ for the $\alpha=0.5$ case is much lower than the $\alpha=0$ case. 
The gap between the inner (red dashed line) and outer (blue dashed line) horizons widens with increasing $\alpha$. This is indicative that the radiation-dominated background for the Kiselev black hole carrying charge increases the gap between the two horizons.

Turning to the modified Painlev\'{e}-Gullstrand setup when $\mathcal{N} \neq 1$, we adopt the square of the shift velocity $v(r)$ in the form

\begin{equation}
 \mathcal{N}^2 = 1+\lambda, \quad   v^2 (r) = (1+\lambda)\left (\frac{2M}{r} - \frac{Q^2-\alpha}{r^2} +\lambda \right )
\end{equation}
where $\lambda$ is an appropriate constant that we restrict to be $\lambda > -1$. Further constraints on $\lambda$ are discussed in the following paragraphs. One sees straightforwardly that the shift function has a maximum located at the point
\begin{equation}
  r_0 = \frac{Q^2 - \alpha}{M}. 
\end{equation}\label{r0_rad}
The corresponding $v(r_0)$ has the value
 
 \begin{equation}
 v(r_0) = \sqrt{(1+\lambda)\left(\frac{M^2}{Q^2-\alpha}+\lambda\right)}=\sqrt{(1+\lambda)\left(\frac{M}{r_0}+\lambda\right)}
 \end{equation}
 
\begin{figure}[!h]
    \centering
    
    \begin{subfigure}{0.45\textwidth}
        \centering
        \includegraphics[width=\linewidth]{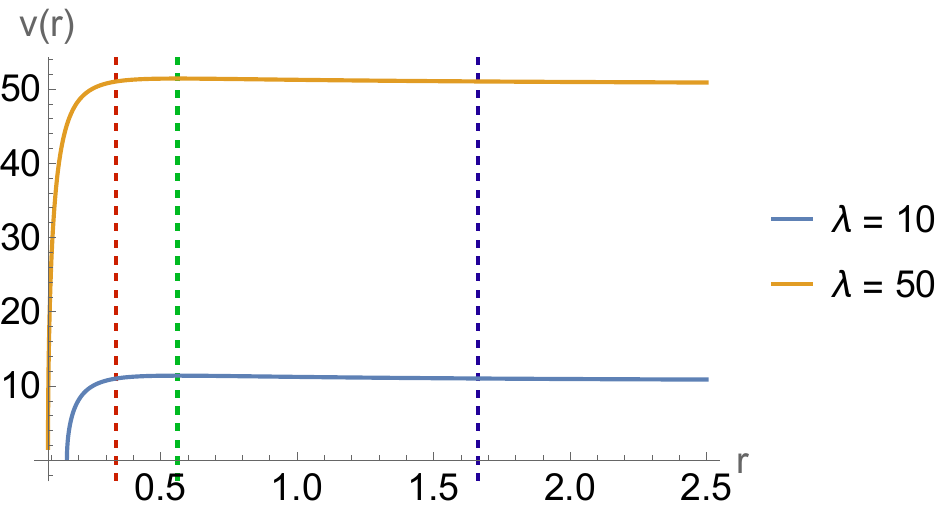}
        \caption{$\alpha=0$}
    \end{subfigure}
    \hfill
    \begin{subfigure}{0.45\textwidth}
        \centering
        \includegraphics[width=\linewidth]{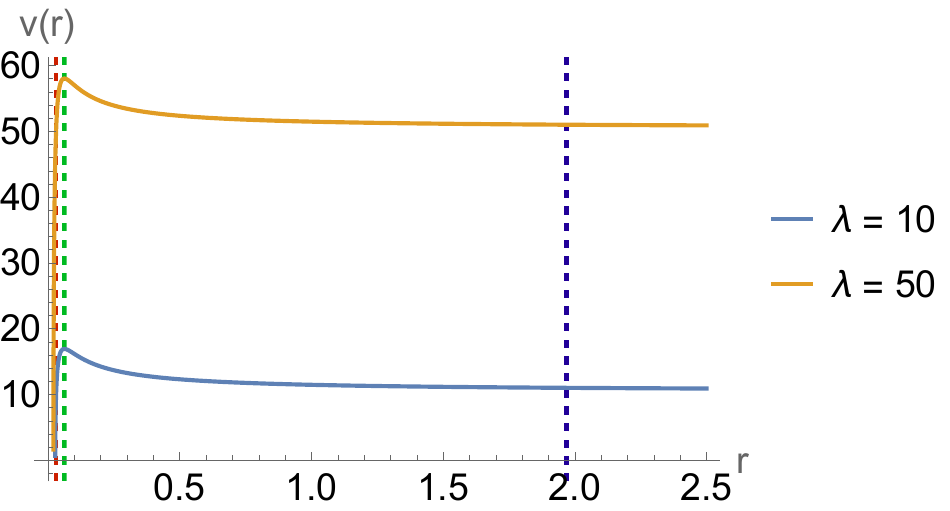}
        \caption{$\alpha=0.5$}
    \end{subfigure}

    \caption{Diagrams illustrating variation of $v(r)$ with $r$ for the radiation case for different values of $\alpha$. We have taken $M=1$ and $Q=0.75$ to exhibit the sub-extremal condition.. The panels correspond to (a) $r_0 = 0.5625$ and (b) $r_0 = 0.0625$. The red dashed line is indicative of the inner horizon $r^{rad}_-$ and the blue dashed line is indicative of the outer horizon $r^{rad}_+$. The green dashed line shows the point $r_0$.}
    \label{v(r)_rad}
\end{figure}

 We find that beyond the maximum point $r_0$, the velocity saturates as the lapse function undergoes changes with respect to the parameter $\lambda$. From the reality of the square root factor of $v(r)$, $\lambda$ is restricted to belong to the region defined by
\[
\big( A \cap B \big) \cup \big( C \cap D \big),
\]
where the individual ones $A, B, C, D$ are such that
\[
A = \{ \lambda \in \mathbb{R} \mid \lambda > -1 \}, 
\quad 
B = \{ \lambda \in \mathbb{R} \mid \lambda > -\frac{M}{r_0} \},
\quad 
C = \{ \lambda \in \mathbb{R} \mid \lambda < -1 \},
\quad 
D = \{ \lambda \in \mathbb{R} \mid \lambda < -\frac{M}{r_0} \}.
\]
Our results showing the nature of $v(r)$ for two different ascending values of $\alpha \in \{0,\, 0.5\}$ are plotted in Fig.\ref{v(r)_rad}.
It points to the specific behavior of $v(r)$ for a range of different values of $\lambda$, each curve tending to saturate at a constant value as $r \rightarrow \infty$. We also observe that with the variation of $\alpha$, the spacing between the inner and outer horizons gradually widens, as can be expected from our previous analysis of $f(r)$, which is related to $v(r)$ via a simple transformation. As $\alpha$ increases, we find that just outside the inner horizon, the velocity attains a sharp peak before eventually decreasing and reaching a constant saturated value.
We also show, for each value of $\alpha$, the velocity trend corresponding to a pair of $\lambda \in \{10,50\}$, including the one that signifies a sudden drop in velocity after the inner horizon is crossed. It is interesting to comment here that Fig.\ref{v(r)_rad}(a), which corresponds to $\alpha=0$, points to the extremal Reisner-Nordstr\"om case for the special condition $M=Q$. Here, the gap between the horizons closes, signifying that both inner and outer horizons merge to a single value. Conditions $A$ and $B$ represent the physical region where, after entering the inner horizon, the particle eventually stops. The higher the value of $\lambda$, the stronger the penetrating power of the particle, eventually reaching a halt position much closer to the singularity at $r=0$. For conditions $C$ and $D$, the behavior is unphysical, in which inside the inner horizon the velocity of the particles suddenly increases exponentially.

\subsection{The dust case}

 The value $\omega = 0$ applies to the situation where the black hole is surrounded by dust. The function $f(r)$ is then given by 

\begin{equation}
     f(r) = 1-\frac{2M + \alpha}{r} + \frac{Q^2}{r^2}
     \label{f_dust_equation}
\end{equation}

\begin{figure}[!h]
\centering

\begin{subfigure}{0.45\textwidth}
    \centering
    \includegraphics[width=\linewidth]{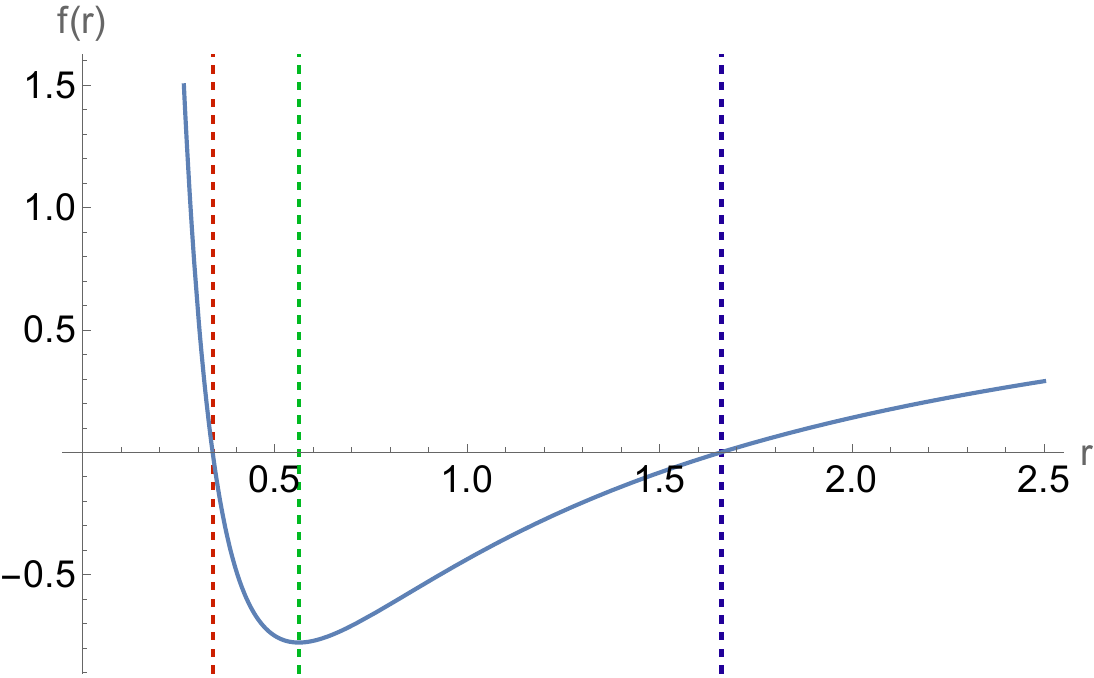}
        \caption{$\alpha=0$}
\end{subfigure}
\hfill
\begin{subfigure}{0.45\textwidth}
    \centering
    \includegraphics[width=\linewidth]{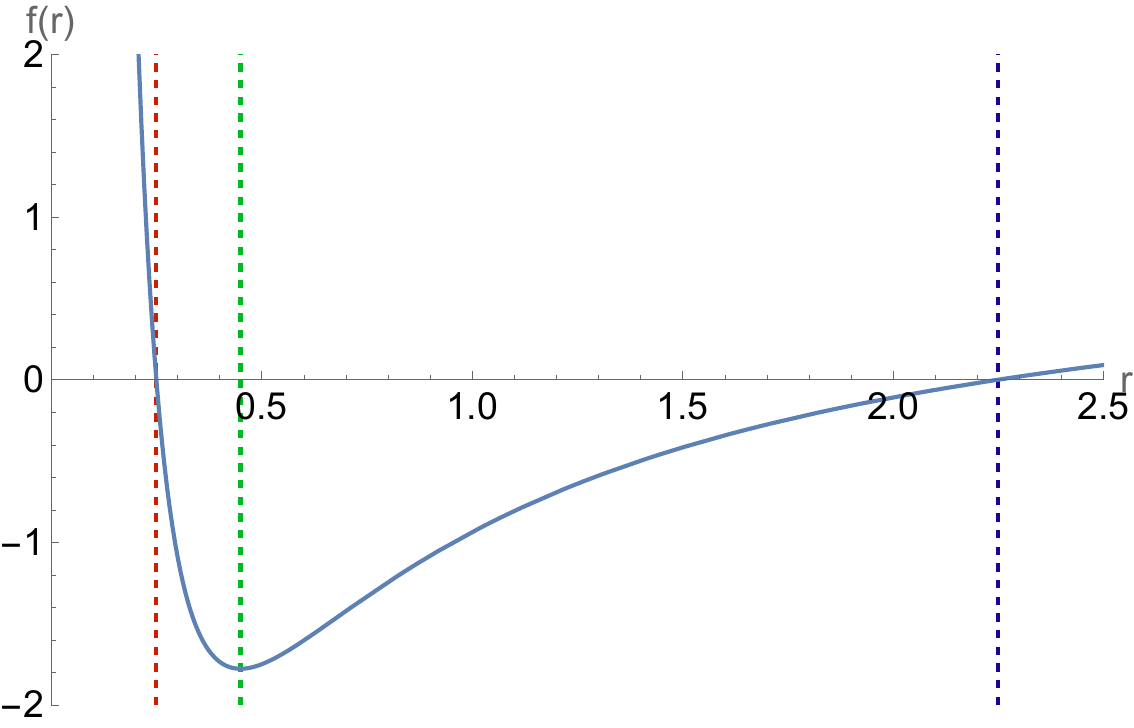}
        \caption{$\alpha=0.5$}
\end{subfigure}

\caption{$f(r)$ vs $r$ for different values of $\alpha$ for the dust case. The blue dashed line represents the outer horizon, the red one represents the inner horizon, and the green dotted line denotes the minimum point $r_0$. We have taken $M=1$ and $Q=0.75$ to show the sub-extremal case.}
\label{f(r)_dust}
\end{figure}

The behavior for different values of $f(r)$ for the dust case is shown in Fig.\ref{f(r)_dust}. The dust case portrays a qualitative behavior similar to that in the radiation case in that the nature of $f(r)$ decreases below a certain value of $r$ by descending to a minimum position and then increases monotonically after undergoing a sign change.

The accompanying square of the shift function acquires the form 
\begin{equation}
     v^2 (r) = \frac{2M +\alpha}{r} - \frac{Q^2}{r^2}
\end{equation}

Computing the following values for the horizon limits from (\ref{f_dust_equation}) by taking $f(r)=0$ we obtain
\begin{equation}
    r^{dust}_{\pm} = \frac{(2M+\alpha) \pm \sqrt{(2M+\alpha)^2-4Q^2}}{2}
    \label{horizon_dust}
\end{equation}
It gives the sum and product of the outer ($r^{dust}_{+}$) and inner ($r^{dust}_{-}$) limits  

\begin{equation}
     r^{dust}_{+} + r^{dust}_{-} = 2(M + \alpha), \quad r^{dust}_{+} r^{dust}_{-}= Q^2 
\end{equation}
which show

\begin{equation}
   r<\frac{r^{dust}_{+}  r^{dust}_{-}}{r^{dust}_{+}+r^{dust}_{-}}
     = \frac{Q^2}{2(M + \alpha)}
\end{equation}
The non-negativity of the square-root term in (\ref{horizon_dust}) places a constraint $\alpha \in ({-\infty,-2(M+|Q|)}]\cup [{-2(M-|Q|),\infty})$. However, only the range $[{-2(M-|Q|),\infty})$ is relevant, since in the other case there are no horizons. \\

Similar to the radiation case, when $\mathcal{N} \neq 1$, we adopt $v(r)$ in the form

\begin{equation}
 \mathcal{N}^2 = 1+\lambda, \quad   v^2 (r) = (1+\lambda)\left (\frac{2M+\alpha}{r} - \frac{Q^2}{r^2} +\lambda \right )
\end{equation}
where $\lambda (> -1)$ is a constant. The shift function $v(r)$ has a maximum at the point $r=r_0$ where the velocity is zero. This gives for $r_0$
\begin{equation}
  r_0 = \frac{2Q^2}{2M + \alpha} 
\end{equation}\label{r0_dust}
The corresponding $v(r_0)$ turns out to be
\begin{equation}
    v(r_0) = \sqrt{(1+\lambda)\left(\frac{(2M+\alpha)^2}{4Q^2}+\lambda\right)} = \sqrt{(1+\lambda)\left(\frac{2M+\alpha}{2r_0}+\lambda  \right)}
\end{equation}
   
\begin{figure}[!h]
    \centering
    
    \begin{subfigure}{0.45\textwidth}
        \centering
        \includegraphics[width=\linewidth]{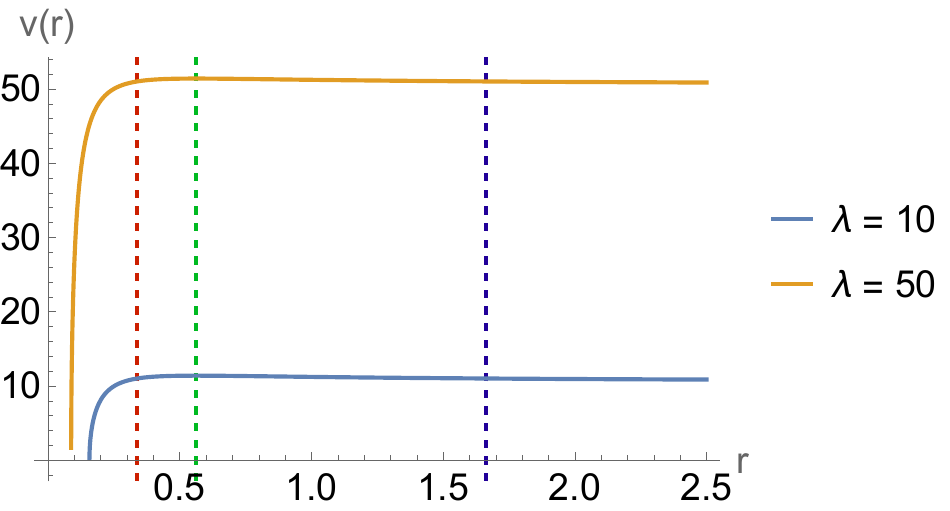}
        \caption{$\alpha=0$}
    \end{subfigure}
    \hfill
    \begin{subfigure}{0.45\textwidth}
        \centering
        \includegraphics[width=\linewidth]{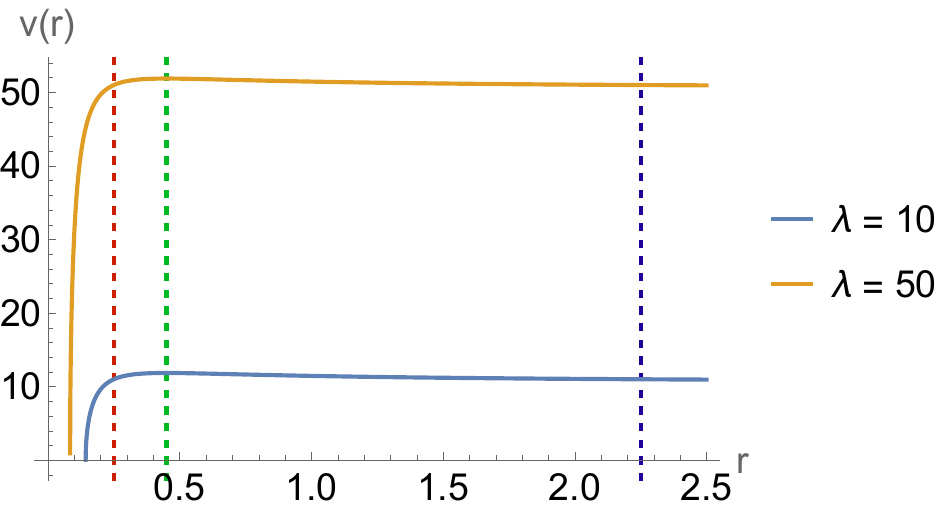}
        \caption{$\alpha=0.1$}
    \end{subfigure}

    \caption{Diagrams illustrating variation of $v(r)$ with $r$ for the dust case for different values of $\alpha$. We have taken $M=1$ and $Q=0.75$. The panels correspond to (a) $r_0 = 0.5625$ and (b) $r_0 = 0.45$. The red dashed line is indicative of the inner horizon $r^{dust}_{-}$ and the blue dashed line is indicative of the outer horizon $r^{dust}_{+}$. The green dashed line shows $r_0$. }
    \label{v(r)_dust}
\end{figure}

In Fig.\ref{v(r)_dust}, for the dust case, we present a schematic plot of $v(r)$ against different values of the quintessence parameter $\alpha$. 
It is at once clear from Fig.\ref{v(r)_dust} that beyond $r_0$ the velocity saturates as the lapse function is varied with respect to the parameter $\lambda$. From $v(r_0)$, we estimate that the restriction on $\lambda$ is given by
\[
\big( P \cap Q \big) \cup \big( R \cap S \big),
\]
where
\[
P = \{ \lambda \in \mathbb{R} \mid \lambda > -1 \}, 
\quad 
Q = \{ \lambda \in \mathbb{R} \mid \lambda > -\frac{2M+\alpha}{2r_0} \},
\quad 
R = \{ \lambda \in \mathbb{R} \mid \lambda < -1 \},
\quad 
S = \{ \lambda \in \mathbb{R} \mid \lambda < -\frac{2M+\alpha}{2r_0} \}.
\]

 The constraints in $\lambda$ are based on conditions $P$ and $Q$, since the other two conditions $R$ and $S$ correspond to the unphysical nature of the velocity profile following a similar argument line as in the radiation case. Note that widening of the gap between the inner and outer horizons is sensitive to increasing values of $\alpha$.

\section{Thermodynamic analysis}

The idea of an event horizon for a black hole essentially came from Bekenstein's works, who used the concept of adiabatic invariants to conjecture \cite{bek1, bek2} that its area was \cite{bek3} quantized according to the square of Planck length $l_P$. With the gradual evaporation of the black hole, transitions happen between discrete mass states with emission of radiation occurring via Hawking radiation \cite{lou}. The laws of black hole thermodynamics were subsequently postulated by drawing a parallel with the standard ones of thermodynamics \cite{hawk1}. Also see an alternate formalism of thermodynamics in terms of quantum tunneling \cite{parikh}, tunneling in BTZ black hole \cite{ssen1}, and tunneling in quintessence driven black hole \cite{ssen2}.  \\

The form for the Hawking temperature can be evaluated from  
\begin{equation}
    T_H=\frac{f'(r_{+})}{4\pi}
\end{equation}
which gives for the respective radiation and dust cases the results

\begin{eqnarray}
    T^{rad}_H &=& \frac{1}{2\pi}\left( \frac{M}{\left(M+\sqrt{M^2-(Q^2-\alpha)}\right)^2}   - \frac{Q^2-\alpha}{\left(M+\sqrt{M^2-(Q^2-\alpha)}\right)^3}  \right) \label{T_rad_equation} \\
T^{dust}_H &=& \frac{1}{\pi}\left( \frac{2M+\alpha}{\left((2M+\alpha)+\sqrt{(2M+\alpha)^2-4Q^2}\right)^2}   - \frac{4Q^2}{\left((2M+\alpha)+\sqrt{(2M+\alpha)^2-4Q^2}\right)^3}  \right)   \label{T_dust_equation} 
\end{eqnarray}

\begin{figure}[!h]
\centering

\includegraphics[width=0.75\textwidth]{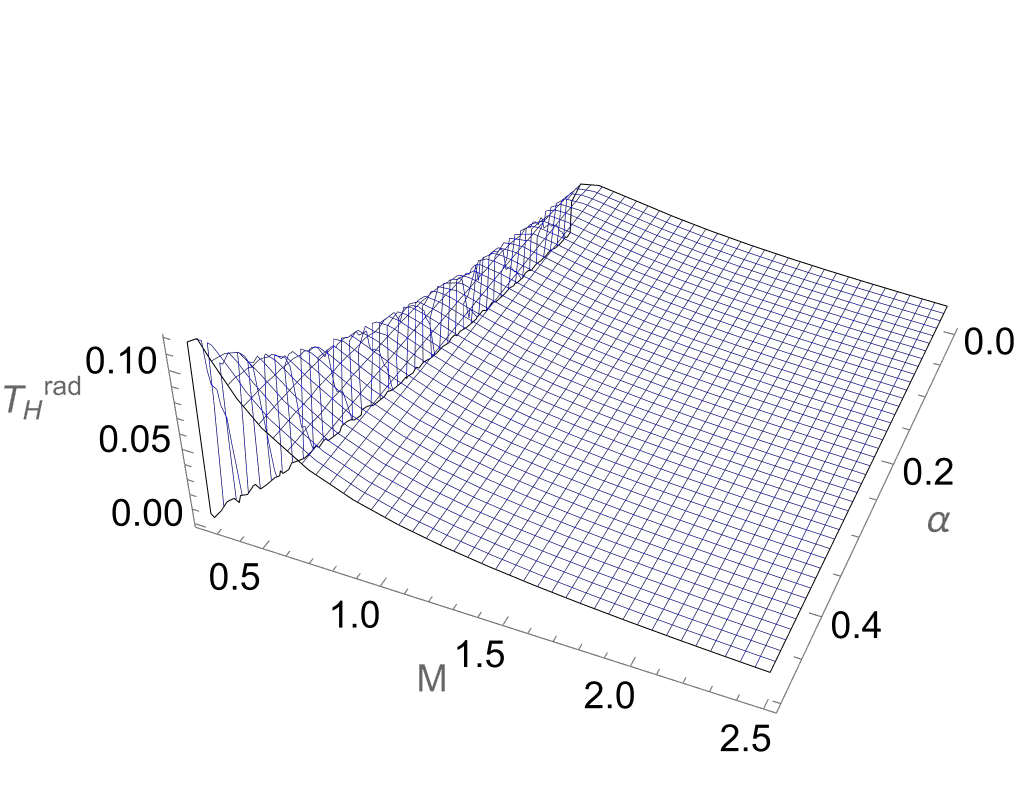}
    \caption{Hawking temperature $T_H$ plotted versus mass $M$ and $\alpha$ for the radiation case. Here we have taken $Q=0.75$. } 
\label{T_rad_mass_surface}
\vspace{1em} 
\end{figure}

\begin{figure}
\begin{subfigure}{0.45\textwidth}
    \centering
    \includegraphics[width=\linewidth]{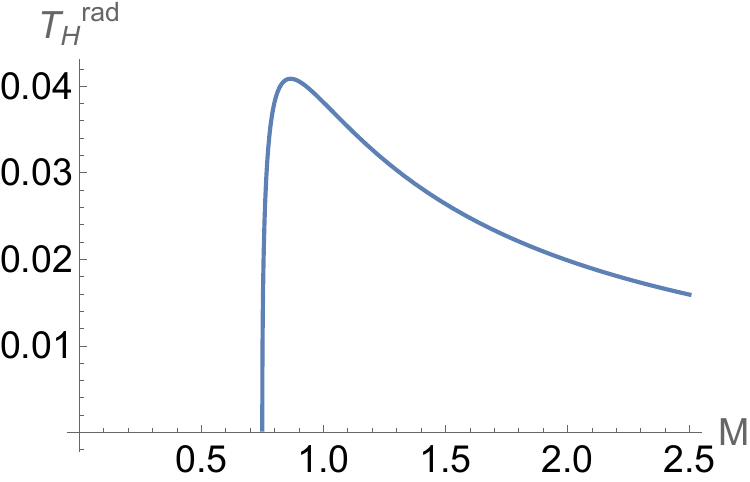}
        \caption{$\alpha=0$}
\end{subfigure}
\hfill
\begin{subfigure}{0.45\textwidth}
    \centering
    \includegraphics[width=\linewidth]{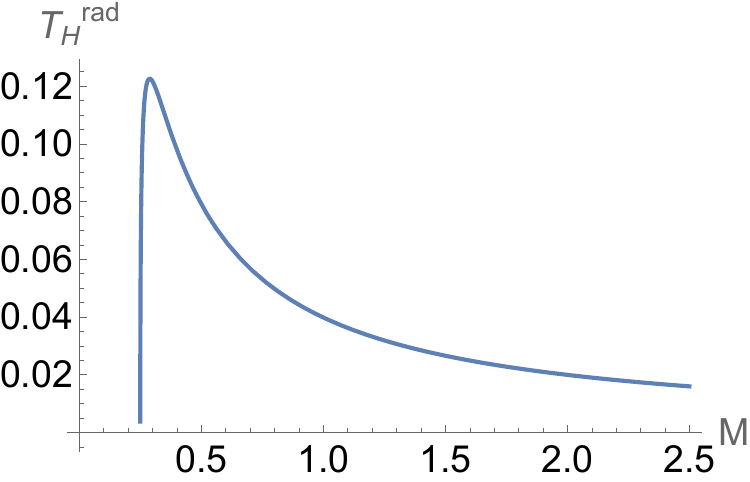}
        \caption{$\alpha=0.5$}
\end{subfigure}

\caption{Hawking temperature $T_H$ vs mass $M$ for different values of $\alpha$ for the radiation case. These plots are the slices of $\alpha=0$ and $\alpha=0.5$ taken from Fig,\ref{T_rad_mass_surface}. Here we have taken $Q=0.75$.}
\label{T_rad_mass}
\end{figure}

\begin{figure}[!h]
\centering

\includegraphics[width=0.75\textwidth]{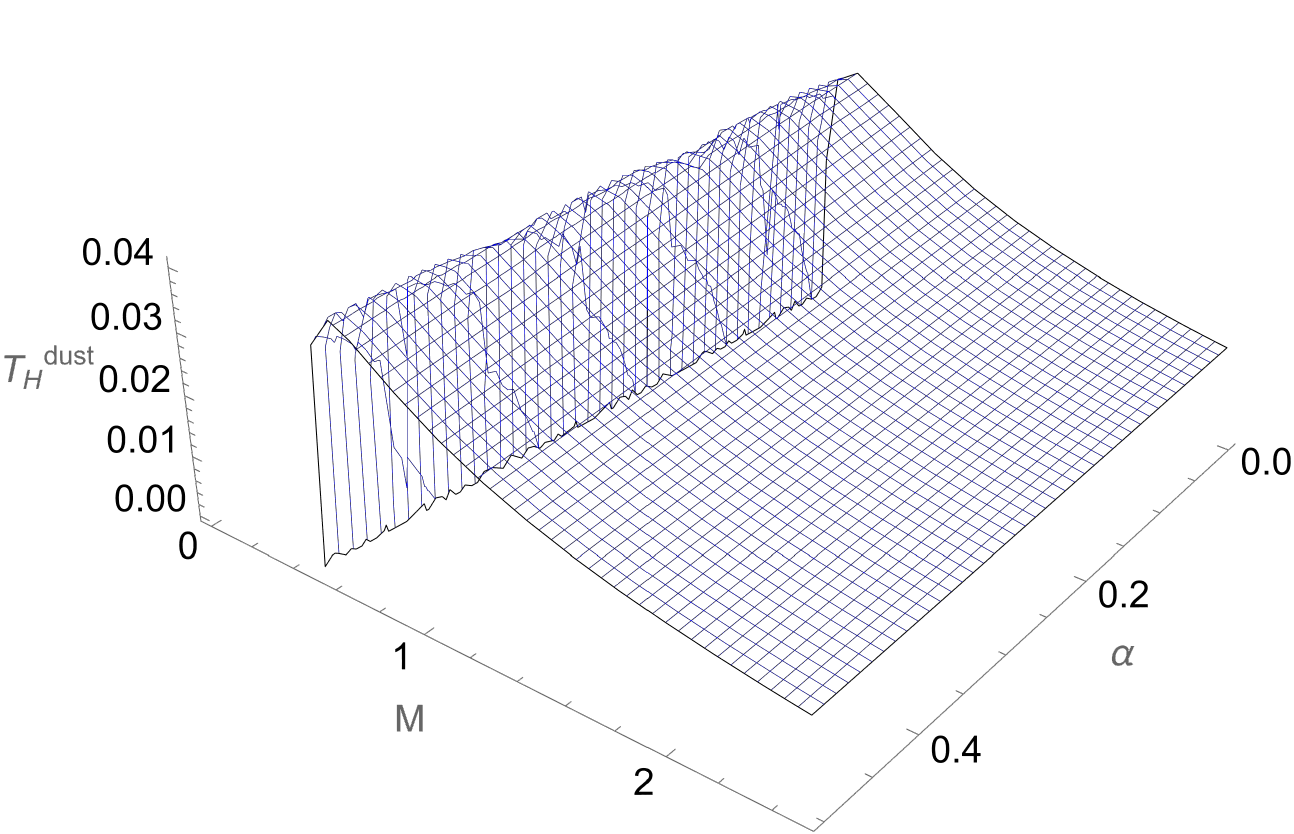}
    \caption{Hawking temperature $T_H$ plotted versus mass $M$ and $\alpha$ for the dust case. Here we have taken $Q=0.75$.}
\label{T_dust_mass_surface}
\vspace{1em} 
\end{figure}

These expressions include the $\alpha$-correction due to quintessence. Its effects are shown in the graphical description of the Hawking temperature versus the mass term $M$ as displayed in Fig.\ref{T_rad_mass_surface} and Fig.\ref{T_dust_mass_surface}. For the specific radiation case, the explicit behavior of $T_H$ with respect to the mass $M$ is plotted in Fig.\ref{T_rad_mass} for different values of $\alpha$. These reveal that with an increase in quintessence $\alpha$, the maximum Hawking temperature value increases. Beyond the peak, the temperature decreases drastically with increasing mass. Note that when the presence of $\alpha$ is turned off, we move on to the familiar Reisner-Nordstr\"om case.

\begin{figure}[!h]
\centering

\includegraphics[width=0.75\textwidth]{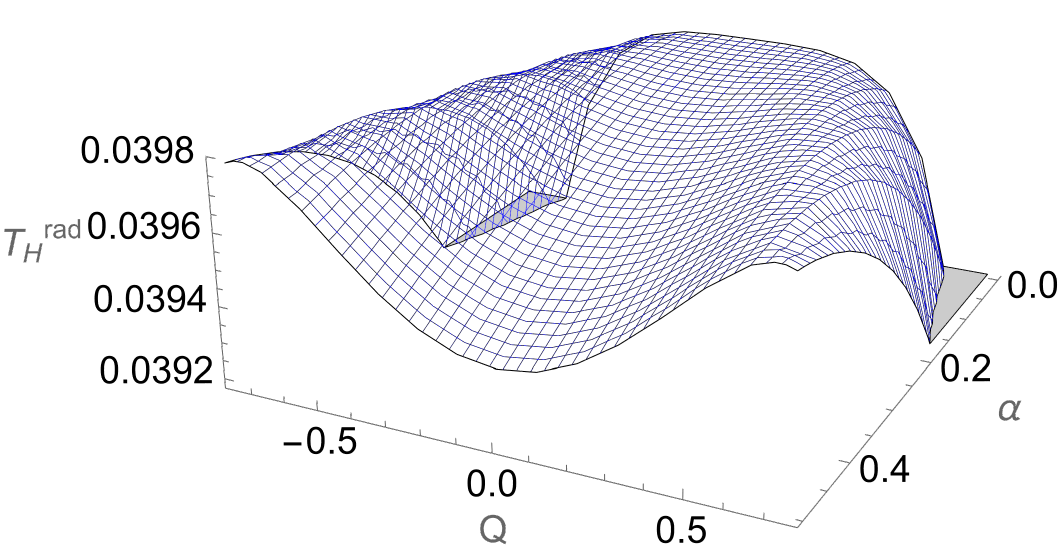}
    \caption{Hawking temperature $T_H$ plotted versus charge $Q$ and $\alpha$ for the radiation case. Here we have taken $M=1$. } 
\label{T_rad_charge_surface}
\vspace{1em} 
\end{figure}

\begin{figure}
\begin{subfigure}{0.45\textwidth}
    \centering
    \includegraphics[width=\linewidth]{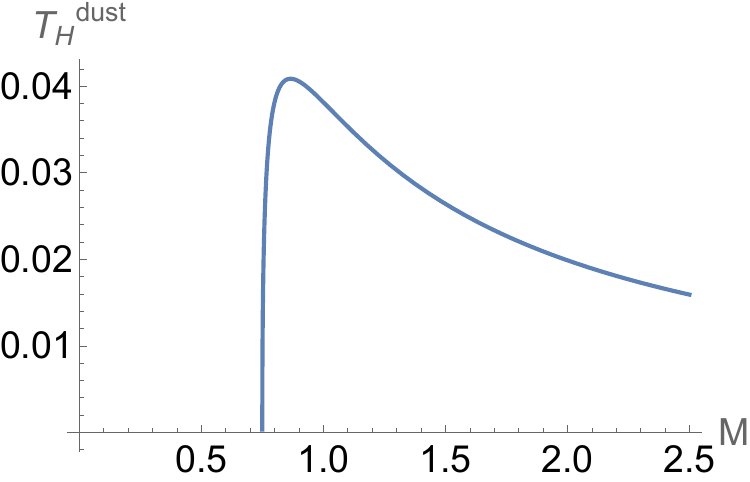}
        \caption{$\alpha=0$}
\end{subfigure}
\hfill
\begin{subfigure}{0.45\textwidth}
    \centering
    \includegraphics[width=\linewidth]{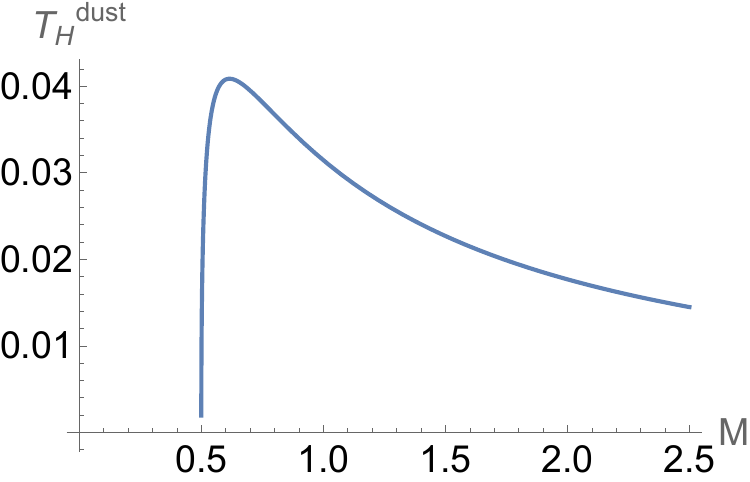}
        \caption{$\alpha=0.5$}
\end{subfigure}

\caption{Hawking temperature $T_H$ vs mass $M$ for different values of $\alpha$ for the dust case. These plots are the slices of $\alpha=0$ and $\alpha=0.5$ taken from Fig.\ref{T_dust_mass_surface}. Here we have taken $Q=0.75$.}
\label{T_dust_mass}
\end{figure}

Fig.\ref{T_rad_mass_surface} shows the variation of the Hawking temperature $T^{rad}_H$ with the black hole mass and the quintessence parameter. The temperature on the outer horizon continues to increase as evaporation continues and the mass $M$ decreases. Eventually, this temperature reaches a peak and suddenly decreases to a zero value. This remnant mass for the radiation scenario at which the temperature goes to zero is the extremal case in which the condition $M = Q^2 - \alpha$ is satisfied. The extremal condition for the dust case is $(2M+\alpha)^2=4Q^2$. This can also be verified for the dust case, as shown in Fig.\ref{T_rad_charge_surface}. Both Fig.\ref{T_rad_mass} and Fig.\ref{T_dust_mass} show the slices of these contour plots for two values of $\alpha$. For the $\alpha=0$ case, when the metric returns to the Reisner-Nordstr\"om case, the peak value $T_H$ is much lower in the radiation case compared to the $\alpha=0.5$ case. This shows that with increasing value of the quintessence parameter $\alpha$, the maximum value of $T_H$ increases monotonically, and the peak temperature for $\alpha \neq 0$ is almost triple for the radiation case. This trend is not very prominent in the dust case. Another observation is that the extremal mass value, where $T_H \rightarrow 0$, is smaller for higher values of $\alpha$ for both radiation and dust cases. 

\begin{figure}
\begin{subfigure}{0.45\textwidth}
    \centering
    \includegraphics[width=\linewidth]{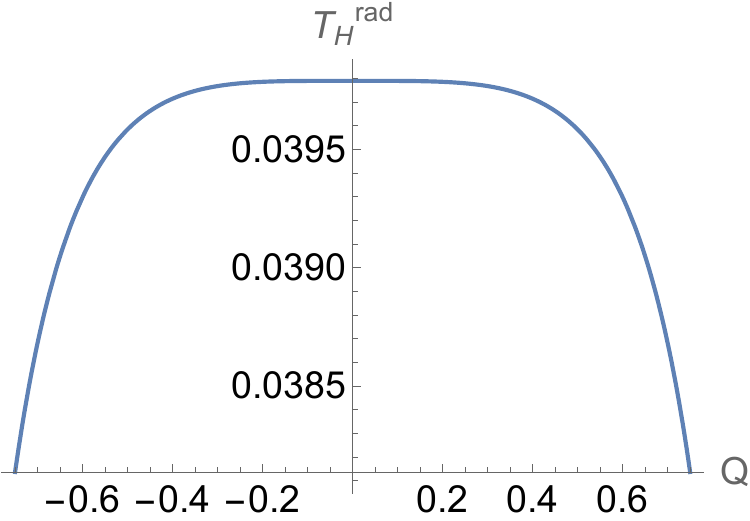}
        \caption{$\alpha=0$}
\end{subfigure}
\hfill
\begin{subfigure}{0.45\textwidth}
    \centering
    \includegraphics[width=\linewidth]{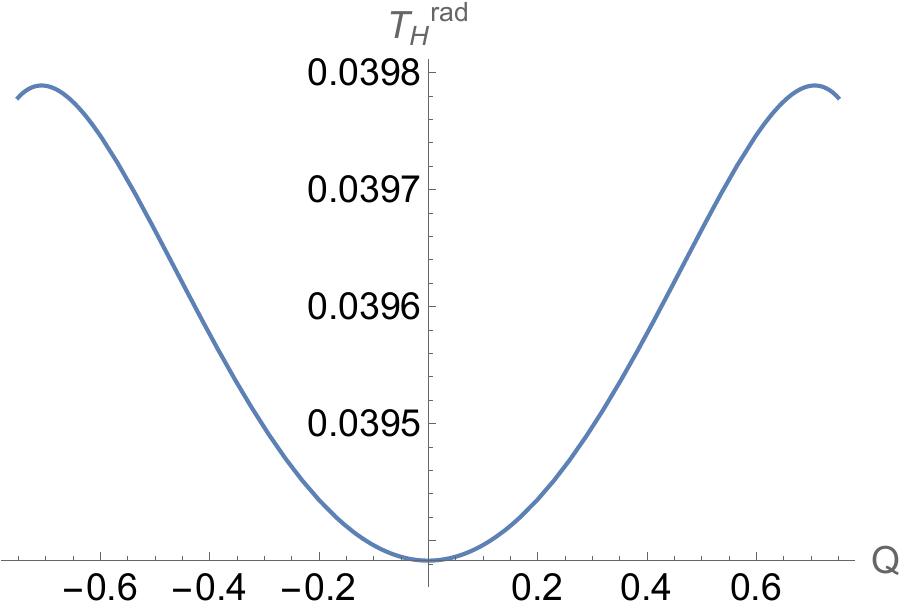}
        \caption{$\alpha=0.5$}
\end{subfigure}

\caption{Hawking temperature $T_H$ vs charge $Q$ for different values of $\alpha$ for the radiation case. These plots are the slices of $\alpha=0$ and $\alpha=0.5$ taken from Fig.\ref{T_rad_charge_surface}. Here we have taken $M=1$.}
\label{T_rad_charge}
\end{figure}

\begin{figure}[!h]
\centering

\includegraphics[width=0.75\textwidth]{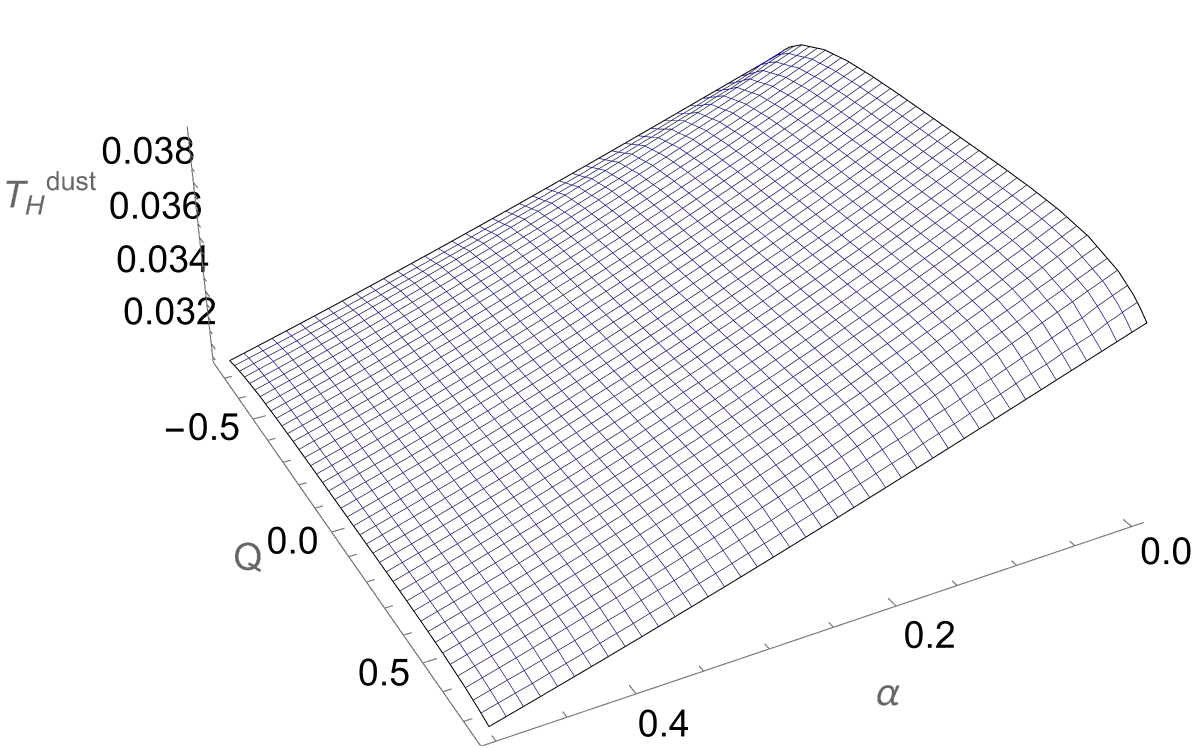}
    \caption{Hawking temperature $T_H$ plotted versus charge $Q$ and $\alpha$ for the dust case. Here we have taken $M=1$.}
\label{T_dust_charge_surface}
\vspace{1em} 
\end{figure}

\begin{figure}
\begin{subfigure}{0.45\textwidth}
    \centering
    \includegraphics[width=\linewidth]{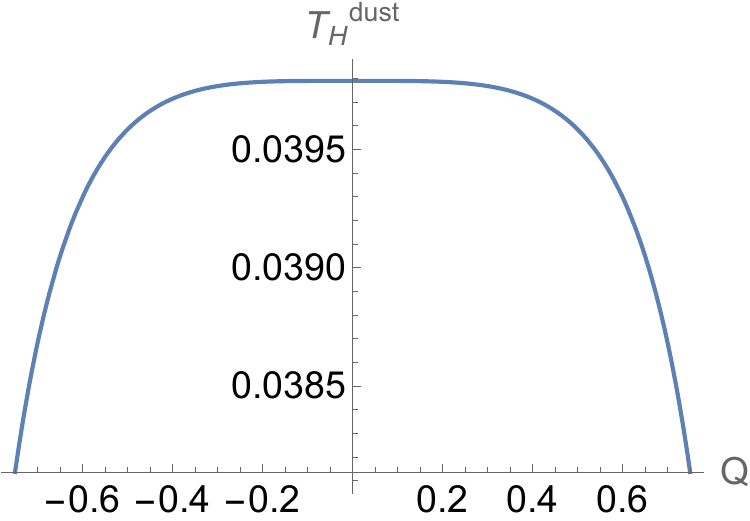}
        \caption{$\alpha=0$}
\end{subfigure}
\hfill
\begin{subfigure}{0.45\textwidth}
    \centering
    \includegraphics[width=\linewidth]{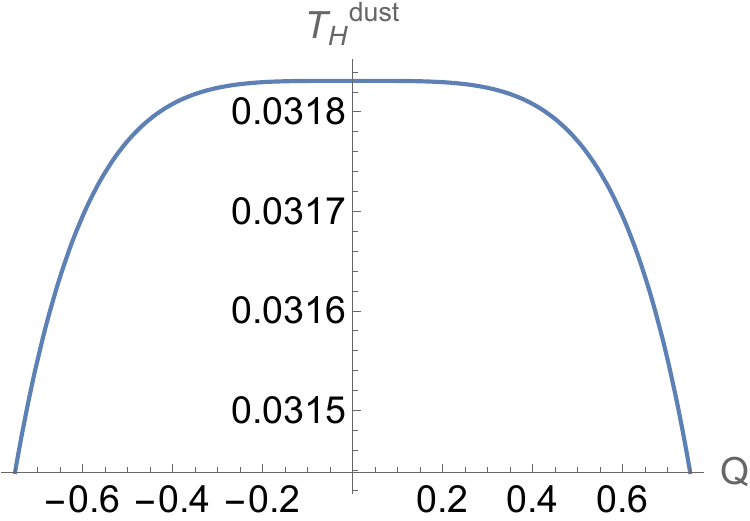}
        \caption{$\alpha=0.5$}
\end{subfigure}

\caption{Hawking temperature $T_H$ vs charge $Q$ for different values of $\alpha$ for the dust case. These plots are the slices of $\alpha=0$ and $\alpha=0.5$ taken from Fig.\ref{T_dust_charge_surface}. Here we have taken $M=1$.}
\label{T_dust_charge}
\end{figure}

When compared to the varying charge, Fig.\ref{T_rad_charge_surface} shows a drastic variation with increasing $\alpha$. Wherein for $\alpha=0$, the value of $T_H$ reaches a maximum around $Q=0$ and a plateau is formed. As $\alpha$ keeps increasing, $T_H$ forms a valley-like dip around $Q=0$, which results in $T_H \rightarrow 0$. . This is evident from Fig.\ref{T_rad_charge} where this feature has been enhanced and portrayed, although the decrease in absolute value of $T_H$ is not much. For the dust case, this trend is not prominent as can be seen from Fig.\ref{T_dust_charge_surface} and Fig.\ref{T_dust_charge}.

\begin{figure}[!h]
\centering

\includegraphics[width=0.75\textwidth]{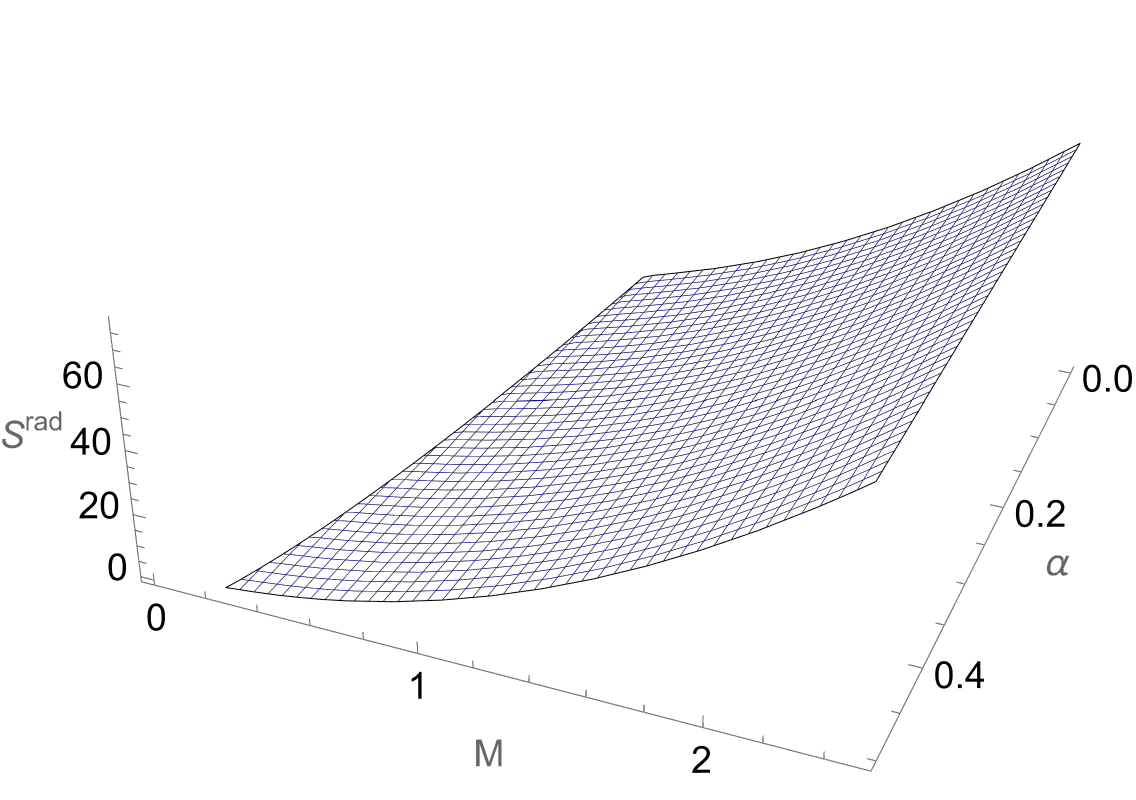}
    \caption{Entropy $S$ plotted versus mass $M$ and $\alpha$ for the radiation case. Here we have taken $Q=0.75$.}
\label{S_rad_mass_surface}
\vspace{1em} 
\end{figure}

\begin{figure}
\begin{subfigure}{0.45\textwidth}
    \centering
    \includegraphics[width=\linewidth]{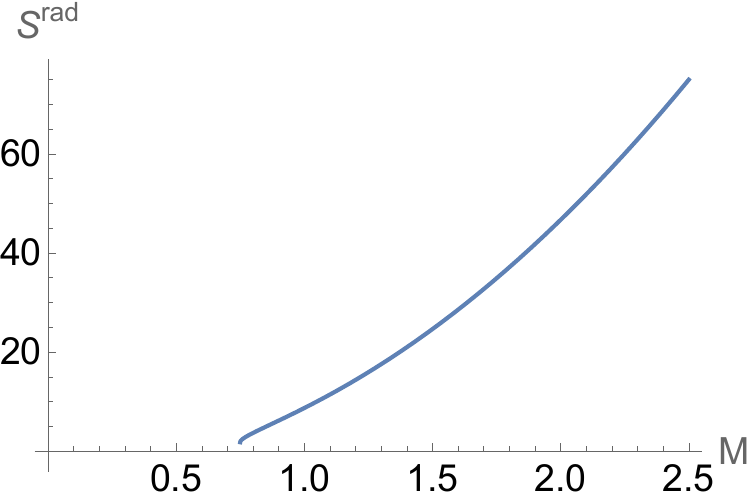}
        \caption{$\alpha=0$}
\end{subfigure}
\hfill
\begin{subfigure}{0.45\textwidth}
    \centering
    \includegraphics[width=\linewidth]{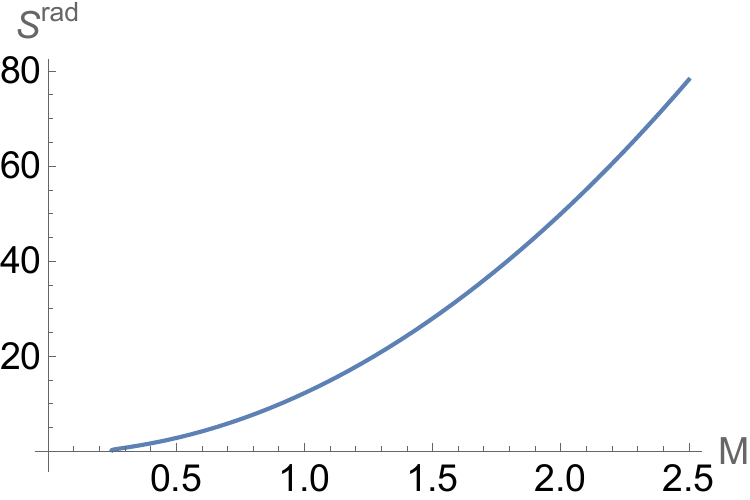}
        \caption{$\alpha=0.5$}
\end{subfigure}

\caption{Entropy $S$ vs mass $M$ for different values of $\alpha$ for the radiation case. These plots are the slices of $\alpha=0$ and $\alpha=0.5$ taken from Fig.\ref{S_rad_mass_surface}. Here we have taken $Q=0.75$.}
\label{S_rad_mass}
\end{figure}

\begin{figure}[!h]
\centering

\includegraphics[width=0.75\textwidth]{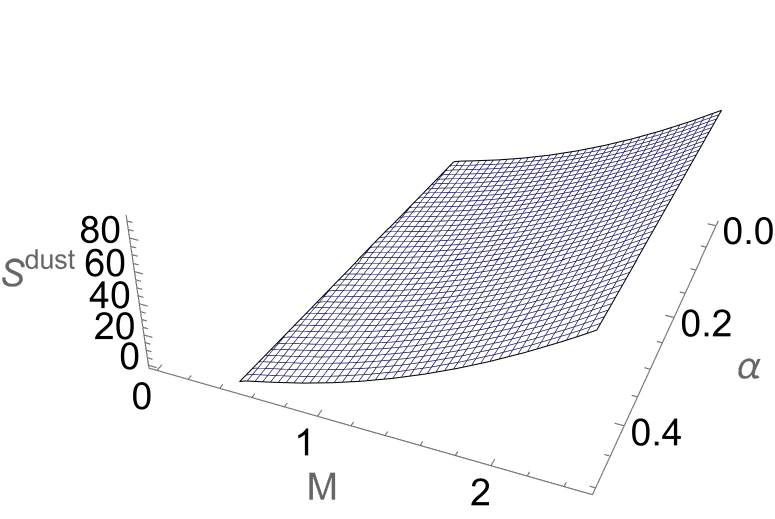}
    \caption{Entropy $S$ plotted versus mass $M$ and $\alpha$ for the dust case. Here we have taken $Q=0.75$.} 
\label{S_dust_mass_surface}
\vspace{1em} 
\end{figure}

\begin{figure}
\begin{subfigure}{0.45\textwidth}
    \centering
    \includegraphics[width=\linewidth]{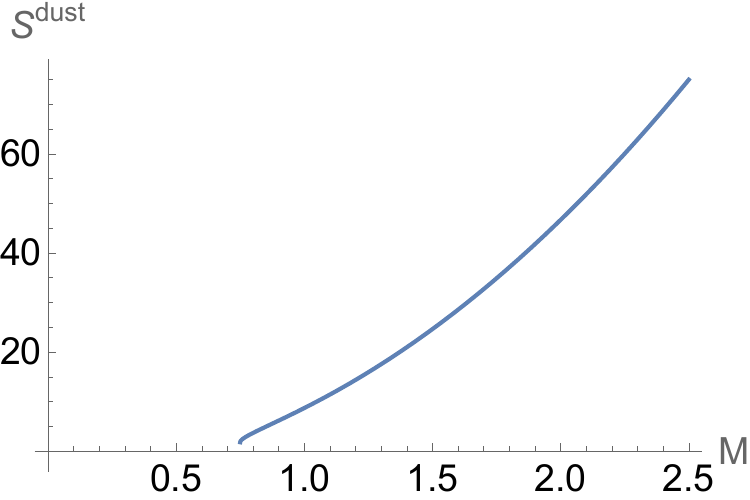}
        \caption{$\alpha=0$}
\end{subfigure}
\hfill
\begin{subfigure}{0.45\textwidth}
    \centering
    \includegraphics[width=\linewidth]{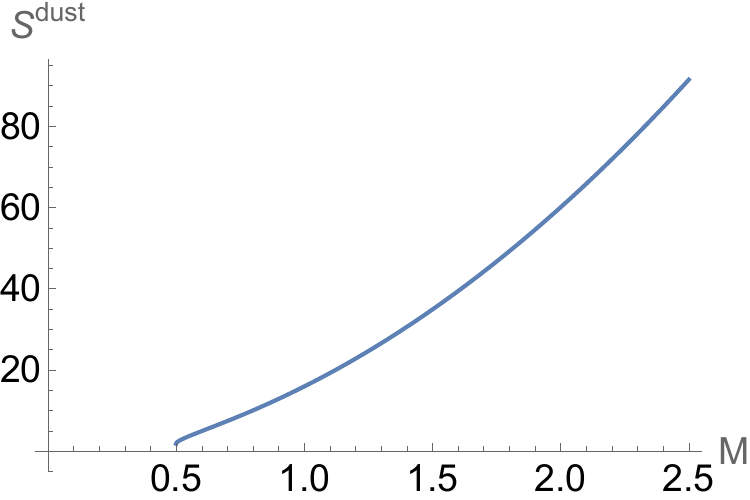}
        \caption{$\alpha=0.5$}
\end{subfigure}

\caption{Entropy $S$ vs mass $M$ for different values of $\alpha$ for the dust case. These plots are the slices of $\alpha=0$ and $\alpha=0.5$ taken from Fig.\ref{S_dust_mass_surface}. Here we have taken $Q=0.75$.}
\label{S_dust_mass}
\end{figure}

The entropy of the black hole is

\begin{equation}
   S=\frac{A}{4} = \pi r_+^2
\end{equation}
where $A=4 \pi r_+^2$. The $\alpha$-corrected forms of the entropy for the Kiselev black hole surrounded by radiation and dust are respectively given by

\begin{eqnarray}
    S^{rad} &=& \pi\left(M+\sqrt{M^2-(Q^2-\alpha)}\right)^2 \label{S_rad_equation}\\
    S^{dust} &=& \frac{\pi}{4}\left((2M+\alpha)+\sqrt{(2M+\alpha)^2-4Q^2}\right)^2
    \label{S_dust_equation}
\end{eqnarray}
The behavior of entropy for the radiation case with respect to the mass $M$ and quintessence parameter $\alpha$ is plotted in Fig.\ref{S_rad_mass_surface} while its monotonically increasing rising behavior is shown in Fig.\ref{S_rad_mass} for different individual cases of $\alpha$-values.  Somewhat analogous results for the dust case are shown in Figs.\ref{S_dust_mass_surface} and Fig.\ref{S_dust_mass}. It is interesting to see that as $Q \rightarrow \sqrt{\alpha}$, the radiation entropy approaches the square of the mass value. For a similar limiting value of $Q$, the dust entropy gets modified with $\alpha$-correction, implying that $S^{dust}$ is influenced by quintessence in contrast to $S^{rad}$.

\begin{figure}[!h]
\centering

\includegraphics[width=0.75\textwidth]{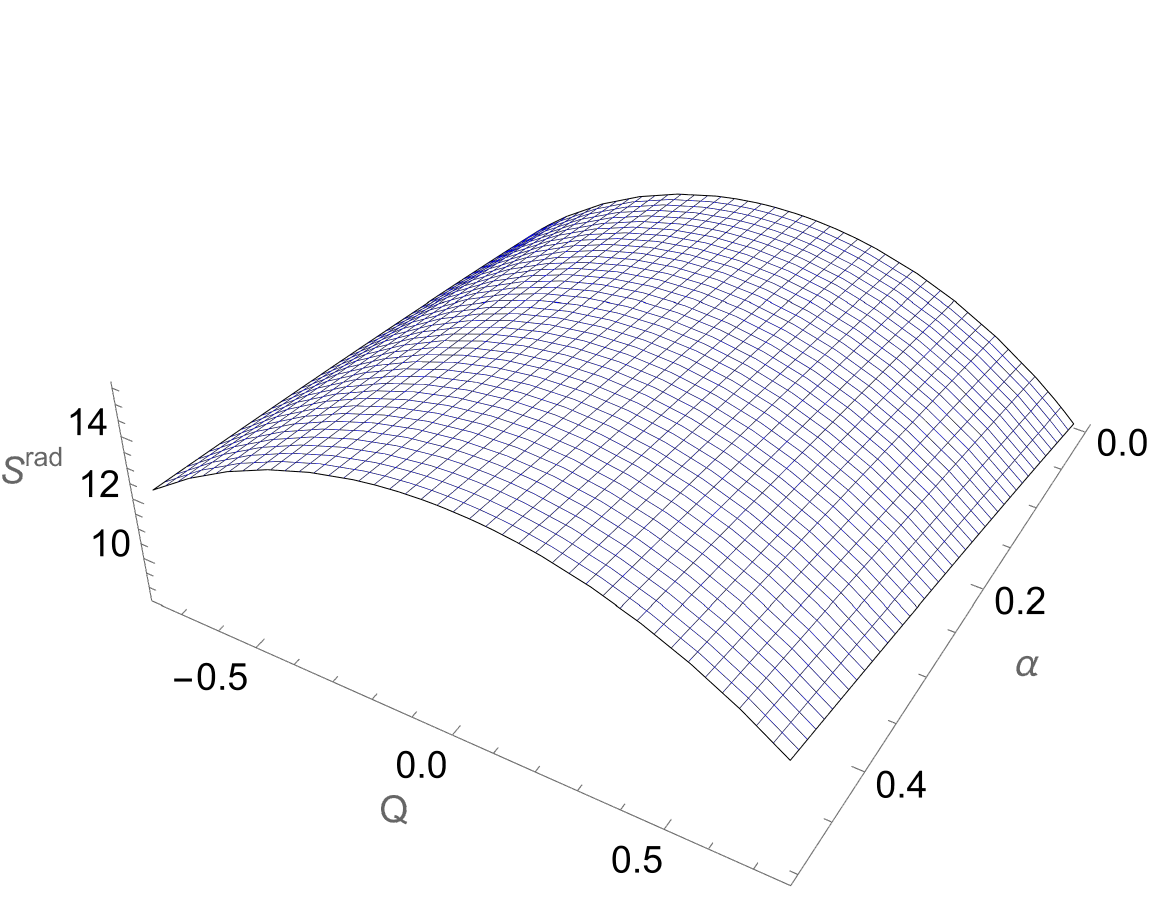}
    \caption{Entropy $S$ plotted versus charge $Q$ and $\alpha$ for the radiation case. Here we have taken $M=1$.}
\label{S_rad_charge_surface}
\vspace{1em} 
\end{figure}

\begin{figure}
\begin{subfigure}{0.45\textwidth}
    \centering
    \includegraphics[width=\linewidth]{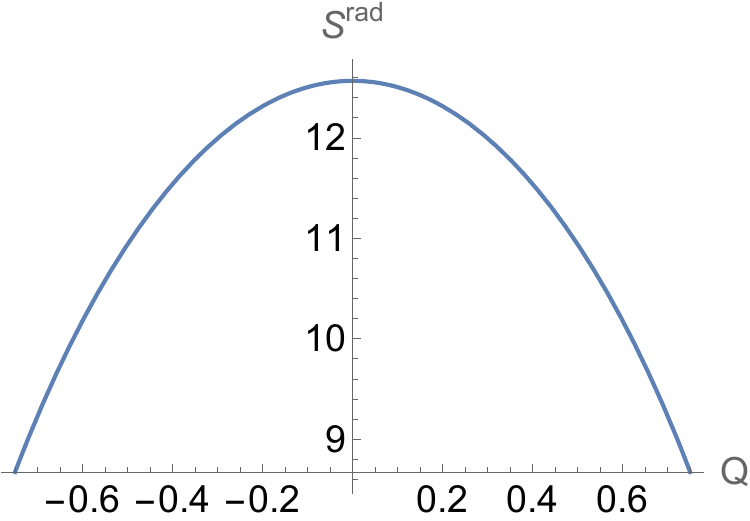}
        \caption{$\alpha=0$}
\end{subfigure}
\hfill
\begin{subfigure}{0.45\textwidth}
    \centering
    \includegraphics[width=\linewidth]{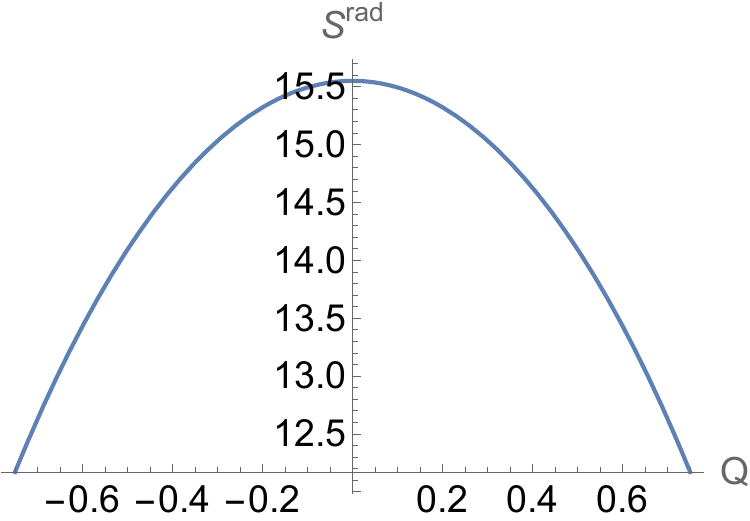}
        \caption{$\alpha=0.5$}
\end{subfigure}

\caption{Entropy $S$ vs charge $Q$ for different values of $\alpha$ for the radiation case. These plots are the slices of $\alpha=0$ and $\alpha=0.5$ taken from Fig.\ref{S_rad_charge_surface}. Here we have taken $M=1$.}
\label{S_rad_charge}
\end{figure}

\begin{figure}[!h]
\centering

\includegraphics[width=0.75\textwidth]{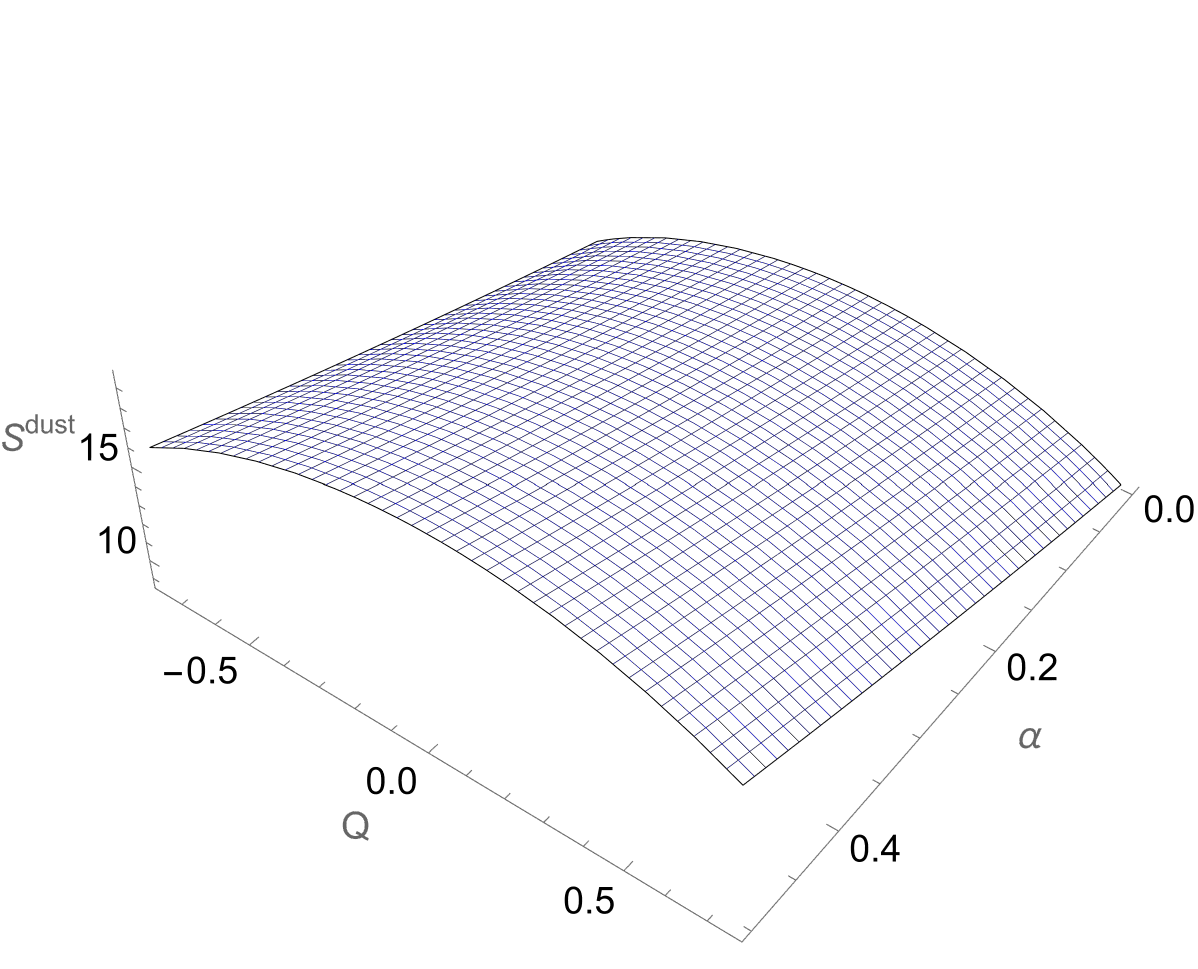}
    \caption{Entropy $S$ plotted versus charge $Q$ and $\alpha$ for the dust case. Here we have taken $M=1$.} 
\label{S_dust_charge_surface}
\vspace{1em} 
\end{figure}

\begin{figure}
\begin{subfigure}{0.45\textwidth}
    \centering
    \includegraphics[width=\linewidth]{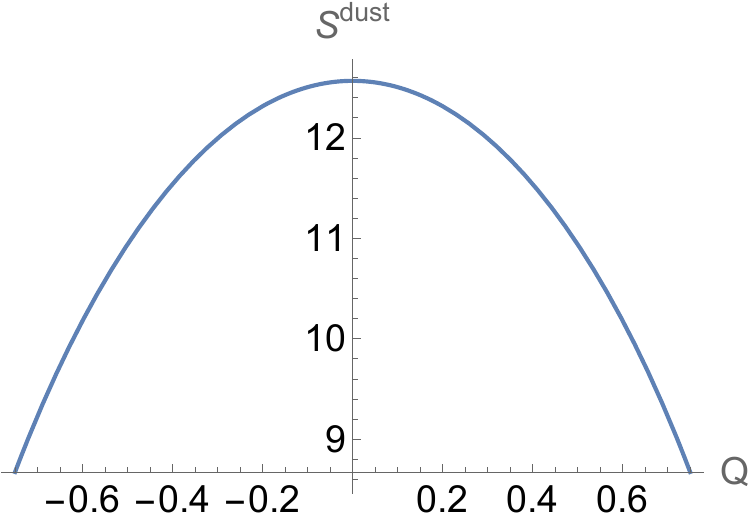}
        \caption{$\alpha=0$}
\end{subfigure}
\hfill
\begin{subfigure}{0.45\textwidth}
    \centering
    \includegraphics[width=\linewidth]{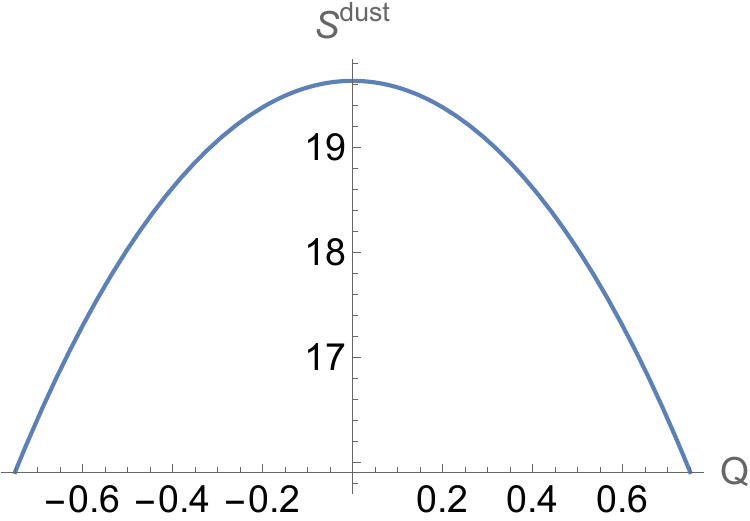}
        \caption{$\alpha=0.5$}
\end{subfigure}

\caption{Entropy $S$ vs charge $Q$ for different values of $\alpha$ for the dust case. These plots are the slices of $\alpha=0$ and $\alpha=0.5$ taken from Fig.\ref{S_dust_charge_surface}. Here we have taken $M=1$.}
\label{S_dust_charge}
\end{figure}

For the radiation case, the variation in entropy follows a uniform and smooth variation with both mass and charge, as can be seen in Fig.\ref{S_rad_mass_surface} and Fig.\ref{S_rad_charge_surface}. The cross sections of these smooth plots for the cases of $\alpha=0$ and $\alpha=0.5$ are shown in Fig.\ref{S_rad_mass} and Fig.\ref{S_rad_charge}. A similar trend can be seen for the dust case as is evident for the variation over mass (Fig.\ref{S_dust_mass_surface} and Fig.\ref{S_dust_mass}) and over charge (Fig.\ref{S_dust_charge_surface} and Fig.\ref{S_dust_charge}).

When the variation in Hawking temperature is observed with respect to mass $M$ for the two cases in Fig.\ref{T_rad_mass} and Fig.\ref{T_dust_mass}, a common feature arises that the mass never goes to zero even when the temperature falls to zero, indicating the remnant mass in the extremal case. But when the Hawking temperature is varied with charge $Q$, it can be seen that for the $\alpha=0$ case, when $Q\rightarrow0$, a peak $T_H$ is achieved. This phenomenon is reversed in the $\alpha=0.5$ value where the $T_H$ is sligtly weaker when $Q\rightarrow0$. This behavior deviates from the uncharged Kiselev black hole case. This peculiar trend is not observed in the dust case, where the $Q \rightarrow 0$ limit gives a residual peak value of $T_H$. 

While analysing the entropy scenario, it is evident for both the radiation and dust cases that entropy always increases with mass $M$ as can be seen from Fig.\ref{S_rad_mass} and Fig.\ref{S_dust_mass} respectively. When varied with charge $Q$, the peak entropy is possible in the uncharged ($Q=0$) case. With increasing charge entropy reduces and reaches a residual value when the square root term vanishes in the expression of entropy for the radiation (\ref{S_rad_equation}) and dust (\ref{S_dust_equation}) case. The same can also be observed through Fig.\ref{S_rad_charge} and Fig.\ref{S_dust_charge}.

\section{Summary}

One of the key advantages of employing Painlev\'{e}-Gullstrand coordinates is that one does not have to encounter coordinate singularities at the event horizon, in contrast to the Schwarzschild metric. In this paper, we have implemented a modified version of Painlev\'{e}-Gullstrand coordinates, defined principally in terms of lapse and shift functions, to investigate certain interesting features of Kiselev black hole spacetime. The metric function of the latter consists of reference to the quintessence parameter $\alpha$ that plays a crucial role in identifying the range of the accompanying shift function through restrictions placed on the limits of the physical horizon arising due to the avoidance of any square-root singularity. In our analysis, we have considered two particular problems of radiation and dust corresponding to the specific values of the barotropic constant. For the radiation case, by exploring the plausible limits on the shift function, we have noticed that the spacing between the inner and outer horizons increases with increasing values of $\alpha$. One such value actually corresponds to the standard Reisner-Nordstr\"om case when the quintessence parameter $\alpha=0$. We observe that increasing the value of the parameter $\lambda$ increases the maximum value where velocity saturates for all cases of $\alpha$. This is true for both the radiation and the dust case. Parameter $\lambda$, which is a signature of the modified Painlev\'{e}-Gullstrand coordinates, is directly proportional to the velocity of the incoming particle falling into the black hole.\\

For the dust case, by performing a similar analysis by plotting $v(r)$ corresponding to different values of the quintessence parameter $\alpha$, we see that beyond the maximum $r_0$, the velocity saturates as the lapse function is varied with respect to the modified Painlev\'{e}-Gullstrand parameter $\lambda$. Specifically, we note that, as in the radiation case, the constraints on $\lambda$ show that the gap between the inner and outer horizons is much wider and enhanced when compared to its dust counterpart. This indicates that Kiselev black hole carrying charge having a radiation-dominated background has a much wider horizon barrier as compared to the dust case.\\

Finally, we made some thermodynamic study of our results by determining the Hawking temperature and entropy for the cases of radiation and dust. The results are obtained in closed forms and show their sensitivity to quintessence. These are graphically illustrated where variation of Hawking temperature is studied with respect to the behavior against charge $Q$, apart from individual variation with $\alpha$ and the black hole mass $M$. We also show explicitly the entropy behavior for the radiation and dust cases in the plots made with respect to the variation in mass $M$ and charge $Q$ for the quintessence parameter $\alpha$. The monotonically rising behavior of the entropy varied against mass $M$ is also noted for different individual values of $\alpha$. As $Q \rightarrow \sqrt{\alpha}$, we observe that the radiation entropy approaches the square of the mass value, while the result for the dust entropy is seen to be influenced by the correction factor arising due to $\alpha$. Another interesting feature to note is that entropy is always maximum for the uncharged Kiselev black hole, whereas there is some remnant mass even when the Hawking temperature $T_H$ goes to zero.

\section*{Acknowledgements}

S.S. acknowledges financial support from the Shiv Nadar Institution of Eminence and the Council of Scientific and Industrial Research (CSIR), Government of India, for a Direct-SRF fellowship under grant  09/1128(18274)/2024-EMR-I.

\section*{Data-availability statement}
All data supporting the findings of this study are included in the article.

\section*{Conflict of interest}
The authors declare no conflict of interest.

\FloatBarrier

\end{document}